\documentclass[aps,prb,twocolumn,superscriptaddress]{revtex4}

\usepackage{graphicx,float}

\begin{document}

\title{Chromium at High Pressures: Weak Coupling and Strong Fluctuations in an Itinerant Antiferromagnet }



\author{R. Jaramillo}
\affiliation{The James Franck Institute and Department of Physics, The University of Chicago, Chicago, Illinois 60637, USA}
\author{Yejun Feng}
\affiliation{The James Franck Institute and Department of Physics, The University of Chicago, Chicago, Illinois 60637, USA}
\affiliation{The Advanced Photon Source, Argonne National Laboratory, Argonne, IL 60439, USA}
\author{J. C. Lang}
\author{Z. Islam}
\affiliation{The Advanced Photon Source, Argonne National Laboratory, Argonne, IL 60439, USA}
\author{G. Srajer}
\affiliation{The Advanced Photon Source, Argonne National Laboratory, Argonne, IL 60439, USA}
\author{H. M. R\o nnow}
\affiliation{Laboratory for Quantum Magnetism, \'{E}cole Polytechnique F\'{e}d\'{e}rale de Lausanne (EPFL), 1015 Lausanne, Switzerland}
\author{P. B. Littlewood}
\affiliation{Cavendish Laboratory, University of Cambridge, Cambridge CB3 0HE, UK}
\author{T. F. Rosenbaum}
\email{tfr@uchicago.edu}
\affiliation{The James Franck Institute and Department of Physics, The University of Chicago, Chicago, Illinois 60637, USA}

\date{\today}

\begin{abstract}
The spin- and charge-density-wave order parameters of the itinerant antiferromagnet chromium are measured directly with non-resonant x-ray diffraction as the system is driven towards its quantum critical point with high pressure using a diamond anvil cell. The exponential decrease of the spin and charge diffraction intensities with pressure confirms the harmonic scaling of spin and charge, while the evolution of the incommensurate ordering vector provides important insight into the difference between pressure and chemical doping as means of driving quantum phase transitions. Measurement of the charge density wave over more than two orders of magnitude of diffraction intensity provides the clearest demonstration to date of a weakly-coupled, BCS-like ground state. Evidence for the coexistence of this weakly-coupled ground state with high-energy excitations and pseudogap formation above the ordering temperature in chromium, the charge-ordered perovskite manganites, and the blue bronzes, among other such systems, raises fundamental questions about the distinctions between weak and strong coupling.
\end{abstract}


\maketitle

\section{Introduction} 

Electron systems are prone to low energy instabilities about the Fermi surface. The most general such instability in the weak-coupling limit is BCS superconductivity~\cite{Tinkham1996}. Formally equivalent to the BCS treatment, but involving solely the spin degrees of freedom, is a short-wavelength magnetic modulation known as a spin density wave (SDW)~\cite{Overhauser1962}. Further coupling between the charge, spin and the lattice may then yield a charge density wave (CDW). These instabilities can coexist and compete, with order parameters that evolve exponentially with the interaction strength.

Although the relationships among the order parameter and its stand-ins (transition temperature, energy gap, superfluid density, magnetic moment) in weak-coupling theory are well established and have been thoroughly tested for thermally-driven transitions~\cite{Tinkham1996}, there have been few if any direct measurements of the exponential dependence of BCS-like order parameters tuned to a quantum critical point at zero temperature. This is particularly important in light of the role played by underlying quantum phase transitions in materials of fundamental interest and potential technological import such as rare earth cuprates exhibiting high-temperature superconductivity~\cite{Coleman2005, Mathur1998, Monthoux2007}, manganites displaying colossal magnetoresistance~\cite{Mannella2005, Yusuf2006, Craco2006}, and transition metal oxides with coupled spin, charge and orbital degrees of freedom at the metal-insulator transition~\cite{Misawa2007}. In each of these cases, the physics involves strong local fluctuations, seemingly at odds with a weak-coupling approach. Yet, the demarcation between weak and strong coupling is not clean cut. It is possible to observe simultaneously seemingly contradictory phenomena in many of these compounds: sliding charge density waves and pseudogaps in the canonical CDW system, the blue bronzes~\cite{Schwartz1995}; spin fluctuations with energies of electron volts and long-coherence-length SDW modulation in the simple elemental antiferromagnet, Cr (Ref.~\onlinecite{Hayden2000}); stripe order on nanometer scales and extended charge density waves in the poster child for local charge fluctuations, the lanthanum manganites~\cite{Milward2005}. 

\begin{figure*}
\begin{center}
\includegraphics{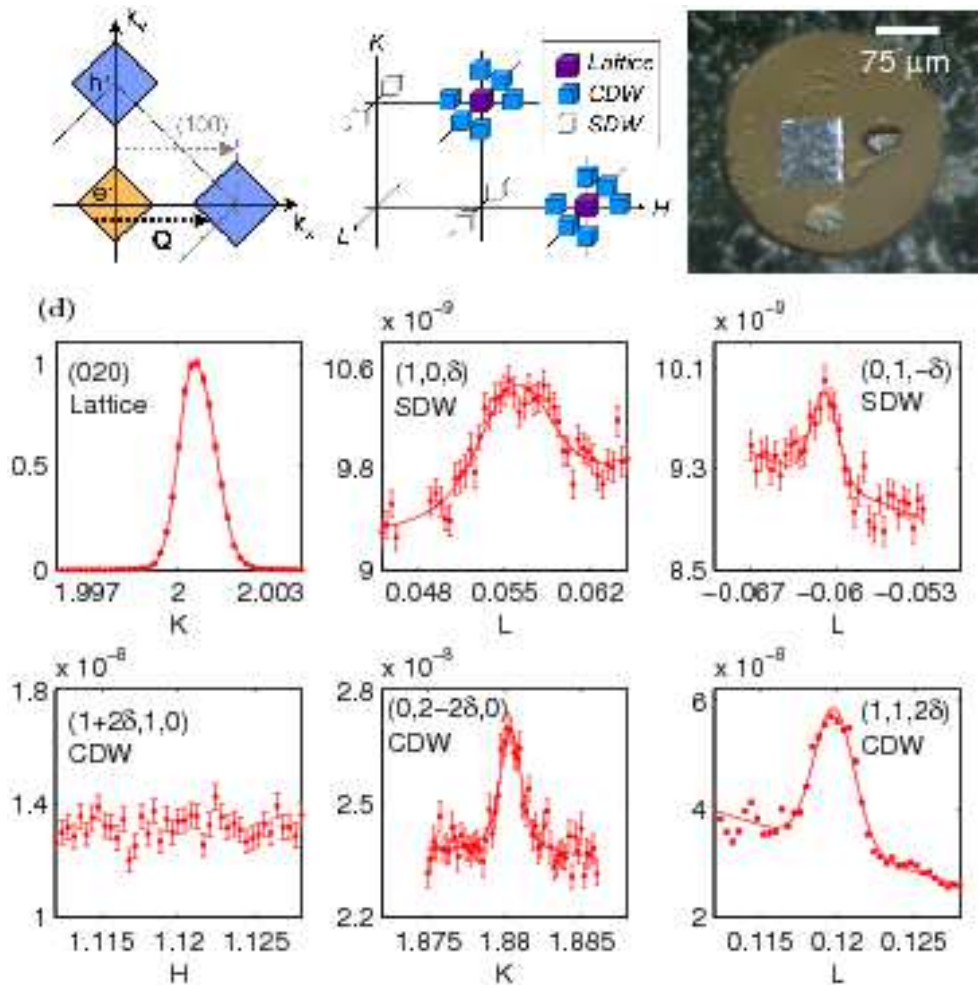}
\end{center}
\caption{\label{FIG_experimental}
Experimental setup and representative diffraction scans. (a) Schematic representation of first Brillouin zone in Cr. Magnetic electron and hole Fermi surfaces are connected by nesting wavevector $Q$; non-magnetic bands are omitted for clarity. (b) Map of satellite CDW and SDW Bragg reflections. CDW peaks appear as a jack around the allowed BCC reflections. SDW peaks appear around the forbidden positions. Only SDW reflections corresponding to a $Q||L$-type domain are shown for clarity. (c) Micrograph of the diamond anvil cell sample chamber showing typical arrangement of an oriented single-crystal Cr sample. Visible as well are Ag and ruby grains that allow pressure measurement \textit{in situ} at low temperature and room temperature, respectively. (d) Scans of SDW, CDW and lattice Bragg peaks at $4.3 \; \mathrm{GPa}$, our highest-pressure SDW measurement. All intensities have been normalized to the value for the nearby ($(0\:2\:0)$ or $(1\:1\:0)$) lattice reflection. The lack of a measurable peak at the $(1+2\delta,\:1,\:0)$ position corresponds to a $\mathbf{Q}||(H\:0\:0)$ volume occupation of below 0.5\%. The difference in width of the two SDW scans is attributed to unequal crystal mosaicities in the different S-domain volumes.
}
\end{figure*}

In this paper, we use hydrostatic pressure to destroy the itinerant antiferromagnetic order in pure chromium metal, providing the clearest demonstration to date of an ordered ground state with an exponentially-tuned (BCS-like) order parameter. A membrane-operated diamond anvil cell held just above liquid helium temperature provides the tuning mechanism, and permits \textit{in situ} measurements of the SDW and CDW order parameters via synchrotron x-ray scattering at the Advanced Photon Source. By pushing the system close to its quantum phase transition, we identify the microscopic terms which couple applied pressure to the ordered magnetic moment. A detailed study of the effects of applied pressure and chemical doping on the magnetic order reveals stark differences between these two means of driving a quantum phase transition. By comparing high temperature transport data for Cr and $\mathrm{Cr}_{1-x}\mathrm{V}_x$ to results for other systems which are typically classified as strongly coupled, namely the stripe-phase manganites and charge-density-wave blue bronzes, we probe the distinction between strong and weak coupling of itinerant electrons and suggest that a hierarchy of energy scales can account for the apparent blurring of these conventional designations. The proximity of our exponentially tuned system to a magnetic instability at high pressure and low temperature highlights the nature of quantum phase transitions for a ground state with no allowed mean-field transition. Our results both confirm some and challenge other long-standing notions on the nature of electronic interactions and instabilities on the Fermi surface. 

Chromium is a 3\textit{d} transition metal with a BCC crystal lattice that has been extensively studied for over forty years as the canonical spin-density-wave system~\cite{Overhauser1962, Fawcett1988}. Its elemental nature relieves complications due to composition which often plague studies of quantum magnetism in other systems and the simple BCC lattice, which undergoes no known structural transition with either pressure or light chemical doping, makes it particularly accessible to conceptual treatment. The itinerant SDW in Cr is stabilized by two nested sheets of Fermi surface, which are eliminated in the magnetic phase by the formation of an exchange-split energy gap~\cite{Fedders1966}. The nesting feature of the paramagnetic Fermi surface, which has been studied by numerical calculation~\cite{Laurent1981} and confirmed by photoemission experiments~\cite{Schafer1999}, results in a quasi-one-dimensional dispersion relation for the magnetic bands in this three-dimensional metal. The SDW is modulated by a wavevector $Q$ (in units of $2\pi/a$, where $a$ is the lattice constant), which is selected by the nesting condition and is slightly incommensurate with the crystal lattice. $Q$ may lie with equal probability along any of the three cubic axes, leading to Q-domains. Below the N\'{e}el temperature, $T_N = 311\;\mathrm{K}$, and above the spin-flop temperature, $T_{SF} = 123 \; \mathrm{K}$, the SDW is transverse and the spins preferentially lie along either cubic axis perpendicular to $Q$, leading to S-domains. Below $T_{SF}$ the SDW is longitudinal. The SDW in Cr is accompanied by an itinerant CDW, which is modulated by $2Q$ and is usually thought of as the second harmonic of the SDW~\cite{Young1974}. This harmonic relationship between spin and charge is consistent with the $I_{CDW} \propto I_{SDW}^2$ scaling (where $I$ is scattering intensity), observed both as a function of temperature~\cite{Hill1995} and pressure~\cite{Feng2007}. 

\begin{figure}[t]
\begin{center}
\includegraphics{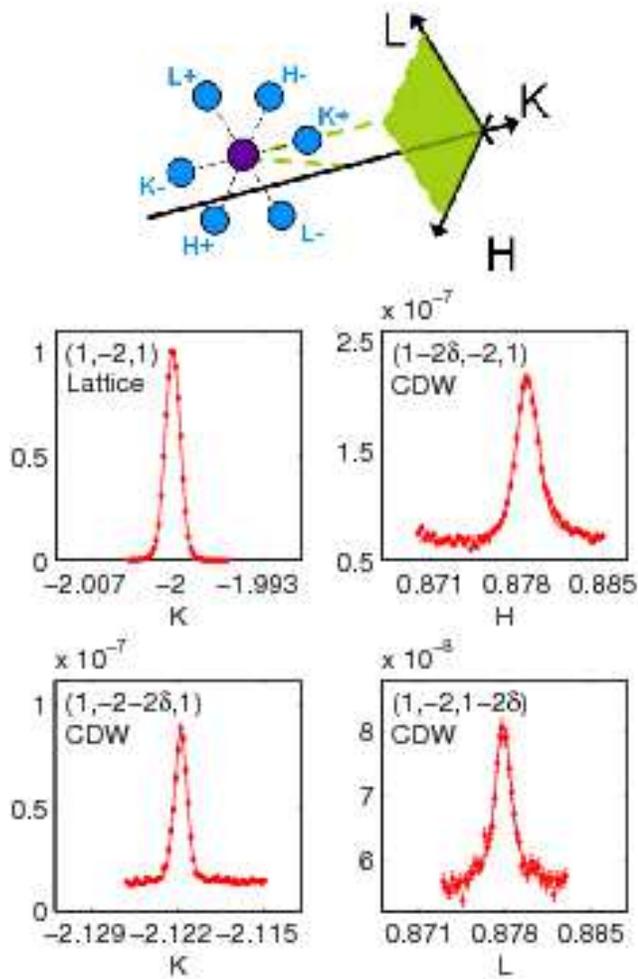}
\end{center}
\caption{\label{FIG_experimental2} 
Lattice and CDW Bragg peaks at $5.0 \; \mathrm{GPa}$; sample oriented with the $[1\:1\:1]$ direction along the diamond anvil cell compression axis. All intensities have been normalized to the value for the $(1\:\bar{2}\:1)$ lattice reflection.
}
\end{figure}

The N\'{e}el transition temperature is suppressed towards zero by applied hydrostatic pressure~\cite{Mitsui1965, McWhan1967, Lee2004, Fuchizaki2006, Feng2007} and/or by sufficient doping with chromium's neighbors in the periodic table~\cite{Werner1967, Yeh2002} (for electron-poor V the critical doping is $\approx 3.4 \%$). By choosing pressure as our means of suppressing the antiferromagnetic state we avoid the effects of disorder and variable electron count that complicate the interpretation of phase transitions driven by chemical doping. In fact, a comparison of these two routes to quantum criticality in Cr illustrates substantial differences in the response of the system to pressure and to doping~\cite{Feng2007}. By directly measuring the spin and charge order parameters as a function of pressure we hope to demonstrate the behavior of this itinerant magnet on a microscopic level, with results that are straightforward to interpret and have the broadest possible relevance to other systems of itinerant electrons with interactions on the Fermi surface.

\section{Experimental Methods}

Direct measurement of the spin and charge density waves was performed using non-resonant monochromatic x-ray diffraction at the insertion device beamline 4-ID-D of the Advanced Photon Source. The capability of probing the spin (SDW) and charge (CDW) order parameters using x-ray diffraction has been demonstrated previously at ambient pressure~\cite{Hill1995}. Here, we extend such measurements to high pressures and liquid helium temperatures~\cite{Feng2007}. The critical pressure at the quantum phase transition exceeds $8 \; \mathrm{GPa}$ (Ref.~\onlinecite{McWhan1967}), necessitating the use of a diamond anvil cell. We employed a home-built, helium-membrane-controlled diamond anvil cell to allow the sample pressure to be changed \textit{in situ} at base temperature with better than $0.05 \; \mathrm{GPa}$ resolution. Pressure was determined \textit{in situ} by measuring the lattice constant of a polycrystalline silver grain included in the pressure chamber volume (Fig.~\ref{FIG_experimental})~\cite{footnote}.

Due to their incommensurate wavevectors, the SDW and CDW Bragg peaks appear as satellites around the forbidden and allowed BCC lattice peaks, respectively. The exponential suppression of the already weak SDW and CDW signals places stringent requirements on the sample quality and instrument collimation. Our samples are miniature Cr single crystals of typical dimensions $(100 \: \times \: 100 \: \times \:40) \; \mu\mathrm{m}^3$ with FWHM from $0.05^{\circ}$ to $0.18^{\circ}$, prepared from a large single-crystal wafer (Alfa Aesar, 99.996+ \%) following procedures in Ref.~\onlinecite{Feng2005}. We have confirmed there is no forbidden lattice peak at the $(1\:0\:0)$ position in our samples even at the highest pressures. We use two different sample cuts, one with $[0\:0\:1]$ along the diamond anvil cell compression axis and another with $[1\:1\:1]$ along this same axis. The first geometry enables measurements of the SDW diffraction satellites around a $(1\:0\:0)$ point (Fig.~\ref{FIG_experimental}), while the second geometry allows the CDW satellites to be measured around a $(1\:\bar{2}\:1)$ point (Fig.~\ref{FIG_experimental2}), which optimizes the structure factors for all three Q-domain types. The use of two separate sample geometries is dictated by the restrictive diffraction geometry of the diamond anvil cell. A Si $(1\:1\:1)$ double-bounce monochromator is used to select $20.000 \; \mathrm{keV}$ x-rays, and a pair of Pd mirrors rejects higher harmonics and focuses the beam to maximize the flux incident on our small sample volume. With the focused high energy monochromatic x-ray beam, highly collimated diffractometer, and 3rd generation synchrotron flux available at 4-ID-D, we achieved a sensitivity of $5\times10^{-10}$ relative to the BCC Bragg intensity (signal $1/10^{\mathrm{th}}$ of background), which is sufficient for following the order parameters into the quantum critical regime.

In order to accurately measure the CDW diffraction intensity, one must account for the Q-domain distribution at each pressure-temperature point. For comparison of the different Q-domain contributions each satellite CDW peak intensity is normalized to the nearest lattice peak intensity (\textit{i.e.} $I(1-2\delta,\: \bar{2},\: 1)/I(1\:\bar{2}\:1)$), taking into account the atomic form factor~\cite{Diana1972} and the geometrical structure factor which expresses the dependence of the cross section on the relative orientations of the scattering wavevector $\mathbf{q}$ and strain wave displacement $\mathbf{u}$ (See Ref.~\onlinecite{Mori1993}). We note that for calculating the Q-domain distribution we only require the angle between $\mathbf{q}$ and $\mathbf{u}||\mathbf{Q}$, and not the actual magnitude of $\mathbf{u}$.

\begin{figure}
\begin{center}
\includegraphics{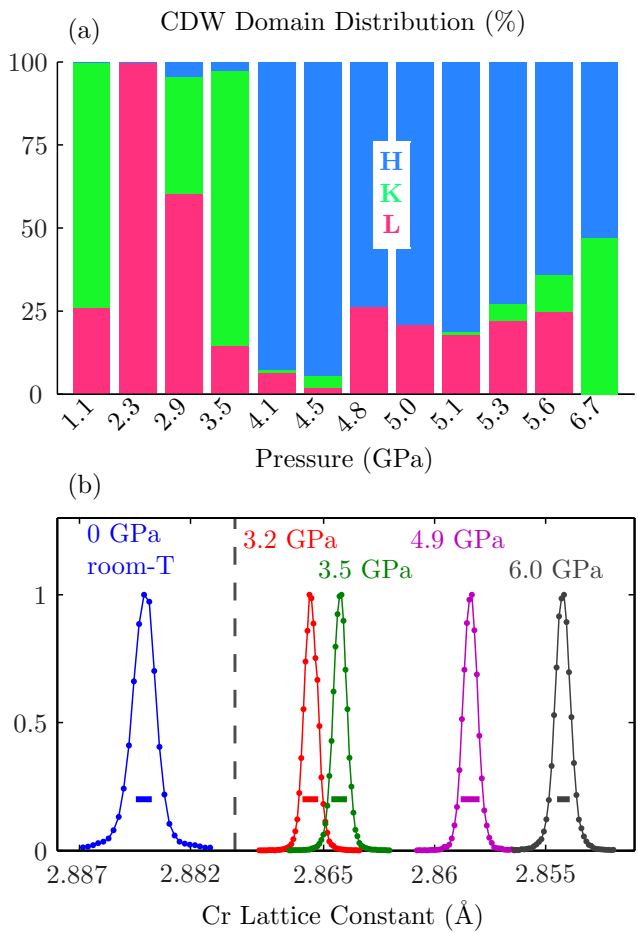}
\end{center}
\caption{\label{FIG_domains-theta2theta}
(a) Evolution of Q-domain distribution with increasing pressure for one Cr crystal at fixed $T = 6.9 \; \mathrm{K}$. The changing domain distribution underscores the need to measure all domain types in order to accurately determine the true amplitude of the order parameters. (b) $\theta-2\theta$  scans of lattice Bragg peaks at a series of pressures; Bragg's law is used to convert from $2\theta$ to Cr lattice constant so that $(1\:1\:0)$ and $(2\:0\:0)$ reflections may be included on the same plot. All scans (except for ambient pressure) are taken at $T < 8 \; \mathrm{K}$. The scalebars associated with each scan give the instrument resolution, taking into account the detector slit resolution, energy resolution and beam divergence (typical values: $70 \; \mathrm{\mu rad}$, $2\times10^{-4}$, and $4 \; \mathrm{\mu rad}$, respectively).
}
\end{figure}

The need to account for the domain distribution at each pressure-temperature point is underscored in Fig.~\ref{FIG_domains-theta2theta}a, where we display the Q-domain distribution for a single sample as the pressure is increased from 1.1 to $6.7 \; \mathrm{GPa}$ while the temperature is maintained at $T = 6.9 \; \mathrm{K}$. The fact that the domain distribution undergoes apparently random changes as pressure is increased also speaks to the quasi-hydrostatic nature of our pressure environment. The crystal represented in Fig.~\ref{FIG_domains-theta2theta}a was oriented in the cell with the L cubic direction along the diamond compression axis. Given the known dependence of the Q-domains on uniaxial stress~\cite{Street1968, Steinitz1970} one would expect that anisotropic stress resulting from the glassy pressure medium would pin the domain configuration into a particular state, most likely along the compression axis. Using our measured value of $(1/a_0)(da/dP) = -1.76\times10^{-3}\;\mathrm{GPa}^{-1}$ for the compressibility of Cr at low temperature and the known value of  $da/a_0 = -17\times10^{-6}$ for the tetragonal strain parallel to $\mathbf{Q}$ at low temperature~\cite{Steinitz1970}, we estimate that the uniaxial stresses affecting our sample are no greater than $0.01 \; \mathrm{GPa}$ whenever any single domain does not occupy 99\% or more of the total volume. 

In addition to a large uniaxial stress affecting the entire sample, the pressure medium might support pressure anisotropies on a smaller length scale. For an estimate of this anisotropy we point to the $\theta-2\theta$ scans in Fig.~\ref{FIG_domains-theta2theta}b. Considering the data at $6.0 \; \mathrm{GPa}$, the measured FWHM is $8\times10^{-4}\;\mathrm{\AA}$ and the calculated instrument resolution is $6\times10^{-4}\;\mathrm{\AA}$. Assuming that this additional broadening provides and upper bound on the pressure anisotropy, and using the measured linear compressibility (above), we calculate an upper bound of $0.1 \; \mathrm{GPa}$. 

Non-resonant magnetic SDW diffraction has an inherently weak cross section
\begin{widetext}
\begin{equation}
\label{EQN1}
\frac{d\sigma_{M}}{d\Omega}=\left(e^2/m_ec^2\right)^2\left(\hbar\omega/m_ec^2\right)^2\left[\left(\mathbf{S_q}\cdot\left(\hat{k}\times\hat{k}'\right)\right)^2+\left(\mathbf{S_q}\cdot\hat{k}\left(1-\hat{k}\cdot\hat{k}'\right)\right)^2\right],
\end{equation}
\end{widetext}
for horizontally polarized x-rays scattered in the vertical plane, where $\hbar\omega$ is the x-ray energy, $\mathbf{S_q}$ is the Fourier transform of the spin distribution evaluated at the momentum transfer $\mathbf{q}$, and $\hat{k}$ and $\hat{k}'$ are unit vectors along the incident and diffracted x-rays, respectively~\cite{Blume1988}. We find that the longitudinal phase is completely suppressed above $\approx 1\;\mathrm{GPa}$ at $8\;\mathrm{K}$, so that all high-pressure measurements presented here are made in the transverse phase. Therefore, barring any accidental equality between the diffraction cross sections for the two types of S-domain that are possible within a given Q-domain, it is necessary to measure two inequivalent SDW reflections (such as $(1,\:0,\:\pm\delta)$ and $(0,\:1,\:\pm\delta)$) in order to determine the S-domain distrubution. The SDW ordered moment is then calculated from the equation
\begin{equation}
\label{EQN2}
\frac{I_{SDW}}{I_{Lattice}}=\left(\hbar\omega/mc^2\right)^2\left(f_m/f\right)\left(\mu/N\right)^2,
\end{equation}
where $f_m$ and $f$ are the magnetic~\cite{Strempfer2000} and atomic~\cite{Diana1972} form factors, $N$ is the number of electrons per site, $\mu$ is the (r.m.s.) ordered moment per atom in units of $\mu_B$, and $I_{SDW}$ and $I_{Lattice}$ are the (properly normalized) SDW and lattice diffraction intensities\cite{Blume1988,Hill1995}. Accounting for the domain structure, we measured $\mu_0\equiv\mu(P=0)=0.39\pm0.02$  at $T = 130\;K$ (above $T_{SF}$), consistent with the accepted value of 0.41 from neutron scattering~\cite{Koehler1966, Fawcett1988}.

The SDW wavevector of single-crystal $\mathrm{Cr}_{1-x}\mathrm{V}_x$ with $x = 3.2\%$ (Ames Lab) was measured under pressure using the triple axis spectrometer TASP at the Swiss Spallation Neutron Source. Pressure was maintained in a neutron compatible compressed helium hydrostatic cell that was mounted in a helium flow cryostat and controlled by an external compressor for \textit{in situ} pressure variation. Pressure was determined \textit{in situ} by measuring the c-axis lattice constant of a pyrolitic graphite crystal which was included in the pressure chamber volume. The lattice constant of this same $\mathrm{Cr}_{0.968}\mathrm{V}_{0.032}$ sample was measured to high resolution at $T = 8 \; \mathrm{K}$ using $17.534 \; \mathrm{keV}$ monochromatic x-ray diffraction at beamline 4-ID-D of the Advanced Photon Source.

Electrical resistivity, $\rho(T)$, measurements of Cr (Alfa-Aesar) and $\mathrm{Cr}_{1-x}\mathrm{V}_x$ (Ames Lab) crystals were performed using a four probe lock-in technique in the Ohmic and low frequency limits in a helium flow cryostat. Samples were cut into rectangular bars and polished before attaching gold leads using a micro spot welding technique~\cite{Walker1998}. For pure Cr two separate samples were used, one having been annealed (20 hr at $1050 \;^{\circ}\mathrm{C}$ in an 85\% Ar, 15\% $\mathrm{H}_2$ atmosphere) to minimize residual lattice strain which is known to affect the shape of the $\rho(T)$ curve near the N\'{e}el temperature.

\section{Results: Tuning the SDW and CDW Ground States with Pressure}

\begin{figure}
\begin{center}
\includegraphics{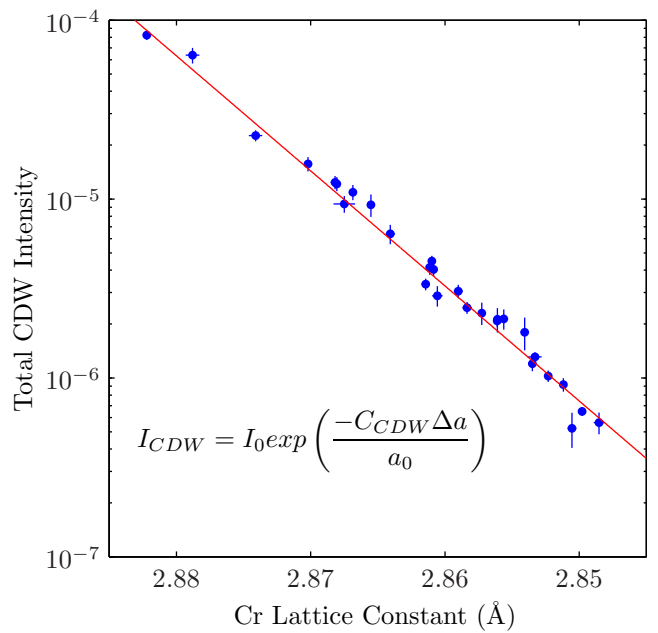}
\end{center}
\caption{\label{FIG_CDW}
Total CDW diffraction intensity $I_{CDW}=I(2Q,\:0,\:0)$ normalized by the lattice reflection $I(2\:0\:0)$ as a function of pressure at $T < 8 \; \mathrm{K}$. Data represent measurements of ten different crystals using two different sample geometries, and are adjusted to account for the measured Q-domain distributions. The intensity decreases exponentially by more than two decades as the lattice shrinks by 1.2\% between 0 and $6.7 \; \mathrm{GPa}$.
}
\end{figure}

We plot in Fig.~\ref{FIG_CDW} the evolution of $I_{CDW}$ at pressures up to $6.7 \; \mathrm{GPa}$ for $T < 8 \; \mathrm{K}$. The quartic relationship between $\mu$ and $I_{CDW}$ enables us to measure over two decades of suppression in $I_{CDW}$ while $T_N$ is decreasing from $311 \; \mathrm{K}$ to $89 \; \mathrm{K}$ (Ref.~\onlinecite{McWhan1967, Feng2007}). This exponential suppression was first demonstrated in our previous work~\cite{Feng2007}; the more complete data set that we present here provides unambiguous proof of this BCS-like ground state. We use the \textit{in situ} low temperature Cr lattice constant rather than pressure as the abscissa because it is determined to better precision for our single crystal samples (Fig.~\ref{FIG_domains-theta2theta}b) and it facilitates comparison of applied pressure with chemical doping (see below). We have
explicitly confirmed that the Cr lattice constant depends linearly on pressure throughout the relevant pressure range~\cite{future}; a linear fit to Cr lattice vs. pressure deviates by less than 1\% from a Birch equation fit at the highest published pressure.

The exponential suppression of the order parameter demonstrated in Fig.~\ref{FIG_CDW} is a general result. Any system of itinerant electrons with an interaction on the Fermi surface will enter an ordered state at low temperature if the interacting vector susceptibility $\chi^{*}(\mathbf{q},\:T)$ diverges at a finite $T$. In a one dimensional metal the non-interacting susceptibility $\chi(\mathbf{q},\:T)$ itself diverges at finite temperature for $\mathbf{q} = 2\mathbf{k_F}$ ($\mathbf{k_F}$ is the Fermi wavevector), and the system undergoes a Peierls transition to a CDW ground state. In higher dimensions $\chi(\mathbf{q}, \: T > 0)$ remains finite, and an ordered state is only possible for a sufficiently strong interaction. The nesting feature of the paramagnetic Fermi surface in Cr results in an enhanced $\chi(\mathbf{q} = \mathbf{Q})$, and the three dimensional electron gas is unusually susceptible to a so-called ``$2k_F$'' transition. The transition in this case is driven by an exchange interaction between nested electron and hole states of opposite spin and results in a SDW ground state~\cite{Overhauser1962}. For an energy gap which is small compared to the Fermi energy, the calculation of the energy gap and mean-field transition temperature is similar to that for a BCS-type superconductor~\cite{Fedders1966}:
\begin{eqnarray}
\label{EQN_BCS1}
& g_0\propto exp(-2\pi^2v/\gamma^2\bar{V}k_c^2) \equiv exp(-1/\lambda) & \\
\label{EQN_BCS2}
& g_0 = 1.76v k_B T_N/\bar{v} \; . &
\end{eqnarray}
$g_{0}=g(T \rightarrow 0)$ is the zero temperature exchange splitting ($2g_{0}$ is the single particle energy gap), $v=\frac{1}{2}(v_a+v_b)$ is the average Fermi velocity for the two nesting bands, $\gamma$ is an average exchange overlap integral, $\bar{V}$ is an average Coulomb potential, $4\pi k_c^2$ is the Fermi surface area of the nesting bands, and $\bar{v}=\sqrt{v_a v_b}$ is the geometric average Fermi volocity. To the extent that the exchange interaction is constant across the nested Fermi surface the ordered magnetic moment $\mu$ is proportional to the energy gap $g$, and using the relationship $\mu\propto I_{SDW}^{1/2} \propto I_{CDW}^{1/4}$ we can track the evolution of the ordered moment and energy gap by measuring the SDW or CDW diffraction intensity. The ground state represented by Eqn.~\ref{EQN_BCS1} is central to much of modern solid state physics, including but not limited to the BCS superconductors, yet experimental verification of this exponential relationship has been lacking. To the best of our knowledge Fig. 4 represents the most convincing demonstration to date of an exponentially tuned BCS-like ground state. 

\begin{figure}
\begin{center}
\includegraphics{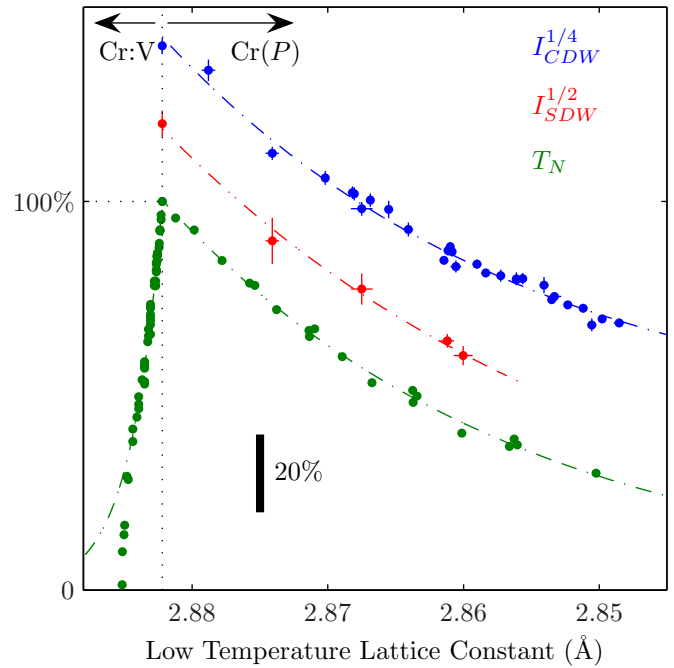}
\end{center}
\caption{\label{FIG_SDW-CDW-TNeel}
Suppression of order parameters and ordering temperatures with applied pressure $P$ (right) and vanadium doping $x$ (left) vs. low temperature lattice constant. Vertical dashed line marks the lattice constant for pure Cr (2.8822 \AA) at ambient $P$ and base $T$. All quantities are normalized to their values for pure Cr at low $T$ and ambient $P$. $I_{SDW}^{1/2}$ and $I_{CDW}^{1/4}$ have been displaced vertically by +0.2 and +0.4, respectively. Diffraction intensities $I_{SDW}$ and $I_{CDW}$ were measured at $T < 8\; K$ and account for the measured Q- and S-domain distributions. $T_{N}(P)$ is taken from Ref.~\onlinecite{McWhan1967} using our measured $(1/a_0)(da/dP)=-1.76\times10^{-3}\;\mathrm{GPa}^{-1}$ and the pressure scale reduced by a factor of 1.3 to account for the difference in calibrations. $T_N(x)$ is taken from Ref.~\onlinecite{CrV, Lee2004} and the lattice constants $a(x,\;T\rightarrow 0)$ are linearly extraloated between the measured values at $x=0$\% and 3.2\% ($2.8850\pm0.0004\;\mathrm{\AA}$). Dashed lines are exponential fits as follows: $I_{CDW}^{1/4}\propto exp\left(\frac{C_{CDW}}{4}\frac{\Delta a}{a}\right)$, $C_{CDW}=436\pm 28$ (68\% c.l.); $I_{SDW}^{1/2}\propto exp\left(\frac{C_{SDW}}{2}\frac{\Delta a}{a}\right)$, $C_{SDW}=227 \pm 10$ (68\% c.l.); $T_N(\mbox{pressure})\propto exp\left(C_{N}^{P}\frac{\Delta a}{a}\right)$, $C_{N}^{P}=110$; $T_N(\mbox{V-doping},\:x<2.5\%)\propto exp\left(C_{N}^{x}\frac{\Delta a}{a}\right)$, $C_{N}^{x}=-1120$. 
}
\end{figure}

The CDW diffraction intensity may be rescaled to demonstrate the $I_{CDW}^{1/4}\propto I_{SDW}^{1/2} \propto \mu \propto T_N$ scaling relationships. We show in Fig.~\ref{FIG_SDW-CDW-TNeel} $\left(I_{CDW}(P)/I_{CDW}(0)\right)^{1/4}$ along with $\left(I_{SDW}(P)/I_{SDW}(0)\right)^{1/2}$ and $T_N(P)/T_N(0)$, where we take $T_N(P)$ from Ref.~\onlinecite{McWhan1967}. The harmonic scaling is confirmed by the exponential fits to the diffraction data, which are in excellent agreement. The ordered moment $\mu$ at the highest reported pressure can be otained from the ratio $(I_{CDW}(P)/I_{CDW}(0))^{1/4} = 0.29 \pm 0.01$ leading to a value of $0.12 \; \mu_B$ at $6.7 \pm 0.1 \; \mathrm{GPa}$, for which the lattice constant is $2.8485 \pm 0.0004 \; \mathrm{\AA}$, a 1.2\% change from ambient pressure. If the applied pressure is resisted primarily by the itinerant electron gas, then a pressure of $6.7 \; \mathrm{GPa}$ corresponds to an increase in energy density of $7.3\times10^{21} \; \mathrm{eV/cm^3}$ for each of the six valence states, or $87 \; \mathrm{meV}$ per valence electron. If we assume that for electrons on the nesting Fermi surface this increase in energy is split evenly between kinetic (band) and potential (exchange) channels, then from these numbers we can estimate the SDW exchange interaction. We define a constant exchange potential $j$ such that the energy required to flip a single ordered spin is $2j\mu$, where $\mu$ is the SDW ordered moment appearing in Eq.~\ref{EQN2}. This is an adaptation of the Heisenberg Hamiltonian $j\mathbf{\mu}_i\cdot\mathbf{\mu}_k$ for a mean-field itinerant magnet, with one moment representing the probe spin that is being flipped, and the other taking on the mean-field value $\mu$. In this way we calculate $j = 0.14 \; \mathrm{eV}$, in agreement with photoemission results on Cr that find a single particle energy gap $2g_0 \approx 0.14 \; \mathrm{eV}$ (Ref.~\onlinecite{Barker1970, Machida1984}). The large energy scale, $j = 140 \; \mathrm{meV}$, dwarfs the relatively small ordering temperature, $T_N(P=0) = 311 \; \mathrm{K} = 26.8 \; \mathrm{meV}$ and is 2\% of the Fermi energy, $E_F = 7.62 \; \mathrm{eV}$ (Ref.~\onlinecite{Laurent1981}).

We plot as well in Fig.~\ref{FIG_SDW-CDW-TNeel} the dependence of $T_N$ on vanadium doping. While doping initially suppresses $T_N$ exponentially, it is markedly different than applied pressure in that the lattice expands (for electron-poor V doping) and the exponential suppression is cut off earlier by a 2nd-order phase transition. The exponential suppression is itself extremely rapid compared to applied pressure, as evidenced by the ratio $\left|C_N^x / C_N^P\right| \approx 10$ of the exponential fit parameters. It was previously shown~\cite{Feng2007} that the suppression of $T_N$ with $x$ and $P$ may be scaled so that the two curves overlap for $x < 2.5\%$; above this point the doped system is driven to a continuous quantum phase transition while the pressurized system remains stable. Such a comparison demonstrates that chemical doping is a faster route to magnetic instability than applied pressure, but by itself does not address the underlying physics of the different responses. 

\begin{figure}
\begin{center}
\includegraphics{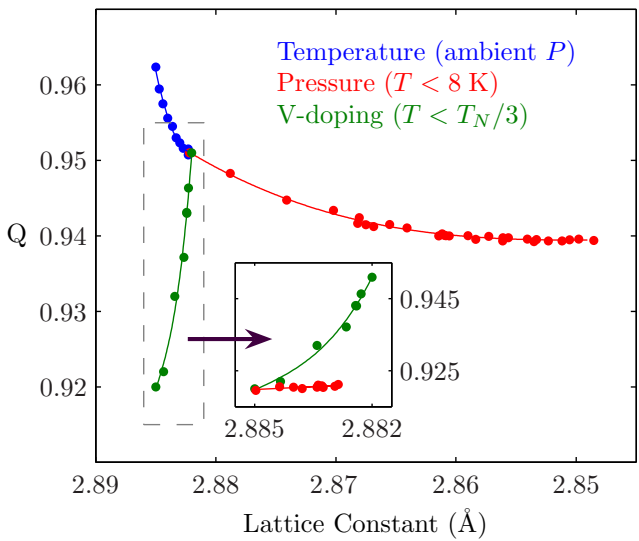}
\end{center}
\caption{\label{FIG_Q}
Evolution of the SDW wavevector $Q$ with pressure, temperature~\cite{Feng2005, Feng2007}, and V-doping~\cite{CrV, Koehler1966}. The rapid variation of $Q$ with $T$ (shown here for $4.2 \; \mathrm{K} < T < 300 \; \mathrm{K}$) reflects the temperature dependence of the Fermi surface as well as the influence of entropy on the magnetic ordering free energy~\cite{Overhauser1962}. The equally rapid variation of $Q$ with doping reflects changes to the Fermi surface resulting from a lowered valence electron count. By contrast, $Q$ in pure Cr varies slowly with $P$, leveling off above $4 \; \mathrm{GPa}$ even as the order parameter continues its exponential decrease (Fig.~\ref{FIG_CDW}). Inset: Evolution of $Q$ with pressure for $\mathrm{Cr}_{1-x}\mathrm{V}_x$, $x = 3.2\%$ at $T < 2 \; \mathrm{K}$ from neutron diffraction. $Q$ is nearly independent of pressure. The highest pressure (smallest lattice constant) plotted here is $4.6 \pm 0.1 \; \mathrm{kbar}$, where $P_C = 7.5 \; \mathrm{kbar}$ (Ref.~\onlinecite{Lee2004}).
}
\end{figure}

Insights into the microscopic mechanisms that drive the suppression of magnetic order can be derived from a consideration of the SDW wavevector $Q$. We present in Fig.~\ref{FIG_Q} a detailed study of $Q$ as a function of temperature, pressure, and chemical doping. At ambient pressure $Q$ decreases rapidly with $T$ (in pure Cr) as the lattice shrinks~\cite{Feng2005}. This decrease in $Q$ results from the diminished importance of entropy to the magnetic ordering free energy, and to the decreasing energetic cost of re-populating reciprocal space: the more tightly $Q$ clamps down on the Fermi surface, the fewer low energy excitations are available, and the more carriers must be re-populated to avoid occupying states above the gap~\cite{Overhauser1962}. The decrease in $Q$ with vanadium doping at low temperature (at ambient $P$) is equally rapid and can be understood as a response of the bandstructure to the reduction in valence electron count on substitution of electron-poor V for Cr~\cite{Schwartzman1989}. By contrast, $Q$ in pure Cr varies slowly under applied pressure at low temperature, even leveling off for $P > 4 \; \mathrm{GPa}$. That $Q$ is constant at high pressure while the order parameter continues its exponential decrease uninterrupted (Fig.~\ref{FIG_CDW}) strongly suggests that the microscopic mechanisms responsible for the suppression of the SDW and CDW intensities cannot be attributed to changes in $Q$. In the inset to Fig.~\ref{FIG_Q}, we ask whether a chemically-doped sample behaves differently by studying the evolution of $Q$ with pressure at low temperature for $\mathrm{Cr}_{0.968}\mathrm{V}_{0.032}$ with $T_N(P=0) = 52 \; \mathrm{K}$. The results from neutron scattering are shown up to $P = 4.6 \pm 0.1 \; \mathrm{kbar}$, more than halfway to the critical pressure of $7.5 \; \mathrm{kbar}$ (Ref.~\onlinecite{Lee2004}). Again, the evolution of $Q$ is clearly pressure independent and contrasts sharply with $Q(x<3.2\%, \: P=0)$ over the same range in lattice constant. 

Our study of $Q(T, \:P,\:x)$ establishes that the magnetic bands in both pure and doped Cr systems are rigid under applied pressure, but are relatively easily deformed by chemical doping. The exponential suppression of magnetic order with applied pressure does not follow from a loss of nested Fermi surface area due to the deformation of the magnetic bands. Rather, it results from an increase in kinetic (band) energy at the expense of potential (exchange) energy, a quantum confinement effect~\cite{Feng2007}. At the same time, the rapid evolution of $Q$ with chemical doping suggests that band structure may in fact play a role in the suppression of magnetic order in the $\mathrm{Cr}_{1-x}\mathrm{V}_x$ series~\cite{Pepin2004}. Given that the exponentially tuned ground state is stable for arbitrarily small values of $\lambda$ (Eq.~\ref{EQN_BCS1}), it will be necessary to follow the data into the quantum critical regime to be able to address the actual nature of the quantum phase transition. 

\begin{figure}
\begin{center}
\includegraphics{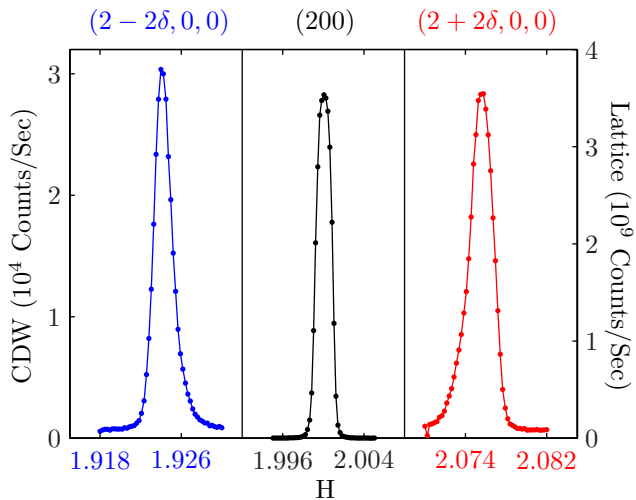}
\end{center}
\caption{\label{FIG_CDW-scans}
Radial scans of CDW satellite pair $(2-2\delta, \: 0, \: 0)$ and $(2+2\delta, \: 0, \: 0)$, and lattice $(2\:0\:0)$ reflection at $295 \; \mathrm{K}$ and ambient pressure; x-axis scaling is the same for all three scans. The asymmetry in the CDW scans is not seen in the $(2\:0\:0)$ reflection, which is sharp and symmetric. The CDW lineshapes are therefore intrinsic to the underlying magnetic order, rather than resulting from the lattice constant distribution. The reflection symmetry of the CDW lineshapes around the $(2\:0\:0)$ position suggests an asymmetric distribution of $Q$ vectors due to deviations of the Fermi surface from idealized flat nesting sheets. 
}
\end{figure}

A high-resolution look at $Q$ reveals subtle deviations of the bandstructure from the idealized nested planes of Fig.~\ref{FIG_experimental}. We plot in Fig.~\ref{FIG_CDW-scans} high resolution scans of the $(2-2\delta, \:0, \:0)$, $(2\:0\:0)$ and $(2+2\delta, \:0, \:0)$ diffraction peaks at ambient pressure and room temperature. The asymmetry in the CDW peaks, which is absent in the lattice peak, suggests an asymmetrical distribution for $Q$ that is unrelated to any residual lattice strain or realistic temperature distribution. The data correspond to a variation in $Q$ of the form $Q = Q_0 + \left|\delta Q\right|$ where $\delta Q/Q_0 \approx 0.1 \%$. As may be expected, this is smaller than the calculated variation of 0.5\% - 1\% in the nesting vector across the magnetic Fermi sheets~\cite{Fawcett1988}. The known deviation of the paramagnetic bandstructure from perfect nesting suggests that the SDW state may accommodate by adopting a distrubtion of wavevectors arising from different regions of Fermi surface; here we display diffraction data with sufficient resolution to support this suggestion. The adaptation of the long-range ordering wavevector to subtle variations in the Fermi surface morphology further emphasizes that the spin density wave in Cr is an electronically soft state.

\section{Discussion: Weak vs. Strong Coupling}

Based on the exponential tuning illustrated in Figs.~\ref{FIG_CDW} and \ref{FIG_SDW-CDW-TNeel} and the soft response of the long range order to the Fermi surface illustrated in Figs.~\ref{FIG_Q} and \ref{FIG_CDW-scans}, it would seem that the SDW in Cr could be characterized definitively as a weakly-coupled ground state. However, there are other phenomena observed in this model system that do not fit into such a cut-and-dry classification. Inelastic neutron scattering has identified spin waves with particularly high velocities (up to $1.5\times10^5 \; \mathrm{m/s}$)~\cite{Fincher1981}, pointing to the presence of a strong magnetic coupling. Magnetic excitations up to $400 \; \mathrm{meV}$ have been observed in nearly antiferromagnetic $\mathrm{Cr}_{0.950}\mathrm{V}_{0.050}$ (Ref.~\onlinecite{Hayden2000}), and our own estimate of $140 \; \mathrm{meV}$ for the exchange interaction would support spin wave modes of up to $280 \; \mathrm{meV}$ if the Heisenberg model is naively invoked. Of particular interest are data that suggest the presence of magnetic interactions above $T_N$. Inelastic neutron scattering intensity from short-range magnetic fluctuations falls off slowly above $T_N$ and is still observed at temperatures above $600 \; \mathrm{K}$ (Ref.~\onlinecite{Fawcett1988}). Measurements of the specific heat and the thermal expansion show clear signatures of incipient order above $T_N$ (Ref.~\onlinecite{White1986, Fawcett1986}). Evidence for high temperature fluctuations is also present in magnetotransport data, with signatures of enhanced scattering and/or loss of carrier density observed in both the Hall and longitudinal resistivities for the $\mathrm{Cr}_{1-x}\mathrm{V}_x$ series~\cite{Yeh2002}. These high temperature signatures of incipient magnetic order and possible pseudogap formation stand in sharp distinction to the canonical weak-coupling theory of spin density waves, for which no magnetic moments exist above the mean-field ordering temperature. Static probes (\textit{d.c.} magnetic susceptibility, elastic scattering) do conform to the expectations of weak-coupling theory; it is the evidence for dynamical non-mean-field effects that resist easy explanation. 

That strong exchange enhancement and dynamical short-range order at high temperature should coexist with a canonical weakly-coupled ground state is not peculiar to Cr. Recent work on the stripe-phase manganites has indicated that the charge stripe phase should be thought of as an itinerant CDW, rather than as a rigid response of the electronic system to the ionic lattice~\cite{Millward2005, Kim2002, Cox2007, Loudon2005, Cox2006}. In particular, the high temperature charge ordered state in $\mathrm{L}_{1-x}\mathrm{Ca}_x\mathrm{MnO}_3$, $\mathrm{L}=(\mathrm{La}, \: \mathrm{Pr})$, for $x > 0.5$ exhibits an ordering wavevector $Q$ which varies continually with temperature and is highly sensitive to lattice strain, both hallmarks of an electronically soft state~\cite{Millward2005, Cox2006}. Other work identifies the charge ordering transition at $T_{CO}$ as a Peierls transition, and evidence has been seen for sliding CDW conductivity~\cite{Cox2007}. This is consistent with the surprising identification of a pseudogap in the canonical sliding CDW system, the blue bronzes~\cite{Schwartz1995}.

\begin{figure*}
\begin{center}
\includegraphics{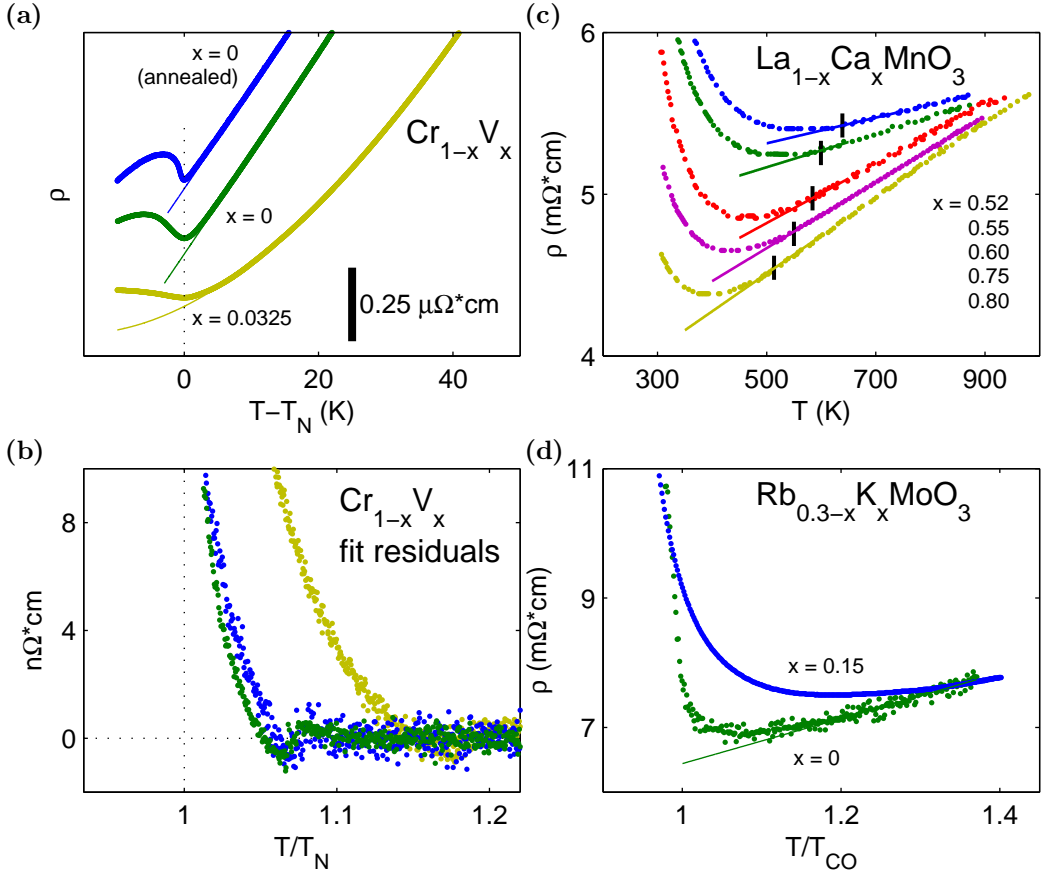}
\end{center}
\caption{\label{FIG_resistivity}
Excess electrical resistivity at high temperature shows effects of incipient long-range order. (a) $\rho(T)$ for pure Cr and $\mathrm{Cr}_{1-x}\mathrm{V}_x$, $x = 3.2\%$ ($T_N =  52 \; \mathrm{K}$). All samples are single crystals; data have been offset vertically to compensate for the very different metallic resistivity at 300 and $50 \; \mathrm{K}$. Cr samples are adjacent cuts from the same wafer, but only one was annealed. Annealing both sharpens the signature at $T_N$ and slightly enhances the excess $\rho$ above $T_N$; therefore this excess $\rho$ cannot be due to lattice strains. Solid lines are cubic fits to the data in the range $T_N + 35 \; \mathrm{K} < T < T_{max}$, where $T_{max} = 400 \; \mathrm{K}$ for pure Cr and $115 \; \mathrm{K}$ for $\mathrm{Cr}_{1-x}\mathrm{V}_x$, $x = 3.2\%$. (b) Excess resistivity, $\rho(T) - \rho_{FIT}(T)$, for the data displayed in (a). The resistivity signature of magnetic fluctuations is seen at a higher normalized temperature in the nearly-critically doped 3.2\% $\mathrm{Cr}_{1-x}\mathrm{V}_x$ than in pure Cr. (c) $\rho(T)$ of a series of $\mathrm{La}_{1-x}\mathrm{Ca}_x\mathrm{MnO}_3$ samples~\cite{Kim2002}. Curves are labeled by doping level $x$ = 0.52, 0.55, 0.60, 0.75, 0.80 for which $T_{CO}$ = 188, 222, 259, 217 and $176 \; \mathrm{K}$, respectively  Black bars mark the pseudogap temperatures $T^*(x)$ determined from optical spectroscopy~\cite{Kim2002}; solid lines are linear fits to $\rho(T,\:x)$ over the range $T > T^*(x$). (d) $\rho(T)$ for pure and doped blue bronze $\mathrm{Rb}_{0.3-x}\mathrm{K}_x\mathrm{MoO}_3$ (Ref.~\onlinecite{Wang2007}); $T_{CO} = 183$ and $179 \; \mathrm{K}$ for $x$ = 0 and 0.15, respectively. Solid line is a linear fit to $\rho(T)$ for $T > T_{CO} + 35 \; \mathrm{K}$.
}
\end{figure*}

We illustrate in Fig.~\ref{FIG_resistivity} the convergence of strong and weak coupling paradigms with high temperature resistivity data for $\mathrm{Cr}_{1-x}\mathrm{V}_x$, $\mathrm{La}_{1-x}\mathrm{Ca}_x\mathrm{MnO}_3$ and $(\mathrm{Rb}_{1-x}\mathrm{K}_x)_{0.3}\mathrm{MoO}_3$. In $\mathrm{Cr}_{1-x}\mathrm{V}_x$ the large excess resistivity below $T_N$ results from a loss of carriers due to the gapped Fermi surface in the SDW phase, while the small excess resistivity seen above $T_N$ (Ref.~\onlinecite{McWhan1967}) suggests the presence of fluctuations associated with incipient magnetic order. Many of the effects associated with fluctuations and reduced effective dimensionality should be less visible to probes with insufficient reciprocal space resolution, such as transport, which are `shorted-out' by the non-magnetic bands. It is therefore noteworthy that such effects are observed in the high temperature resistivity of $\mathrm{Cr}_{1-x}\mathrm{V}_x$. In $\mathrm{La}_{1-x}\mathrm{Ca}_x\mathrm{MnO}_3$, the significant excess resistivity in the range $T_{CO} < T < T^*$ shows the influence of a strong pseudogap on transport, and the energy scale $T^* > 2T_{CO}$ speaks to the strongly correlated nature of the manganites. Nevertheless, below $T_{CO}$ the $\mathrm{La}_{1-x}\mathrm{Ca}_x\mathrm{MnO}_3$, $x > 0.5$ series enters into a CDW state for which the BCS-like description should be applicable and where, in fact, the evolution of the energy gap below $T_{CO}$ obeys the mean-field form for $x \geq 0.60$ (Ref.~\onlinecite{Kim2002}). Analogous behavior is seen in quasi-one-dimensional CDW systems such as the blue bronzes, $(\mathrm{Rb}_{1-x}\mathrm{K}_x)_{0.3}\mathrm{MoO}_3$ (Ref.~\onlinecite{Schwartz1995, Wang2007}), for which fluctuations dominate over a large temperature range $T_{3D} < T < T_{MF}$. Here $T_{MF}$ is the predicted mean-field transition temperature for the one-dimensional electron gas and $T_{3D}$ is the observed long-range ordering temperature.

Further evidence for an interesting high-temperature regime in Cr comes from considering the relative energy scales $g_0$ and $k_BT_N$. BCS theory predicts $2g_0 = 3.5k_BT_N$. Although this relationship is altered for Cr by the fact that multiple bands are involved and the appropriate relationship is given instead by Eq.~\ref{EQN_BCS2}, the Fermi velocities of the two magnetic bands differ by only $\approx15\%$ (Ref.~\onlinecite{Laurent1981}) and the theoretical correction to the BCS value of 3.5 is less than 1\%. Experimentally, however, optical measurements~\cite{Barker1970} of a series $\mathrm{Cr}_{1-x}\mathrm{M}_x$, $\mathrm{M}=(\mathrm{V}, \: \mathrm{Ru}, \: \mathrm{Mn})$ including pure Cr have shown $2g_0 = 5.1k_BT_N$. This suppression of the long-range ordering temperature below the value expected from mean-field theory points to the effect of fluctuations, and is consistent with the mismatch between the larger energy scales present in the system (such as the calculated $j = 140 \; \mathrm{meV}$) and the observed $T_N$. 

\begin{figure}[t]
\begin{center}
\includegraphics{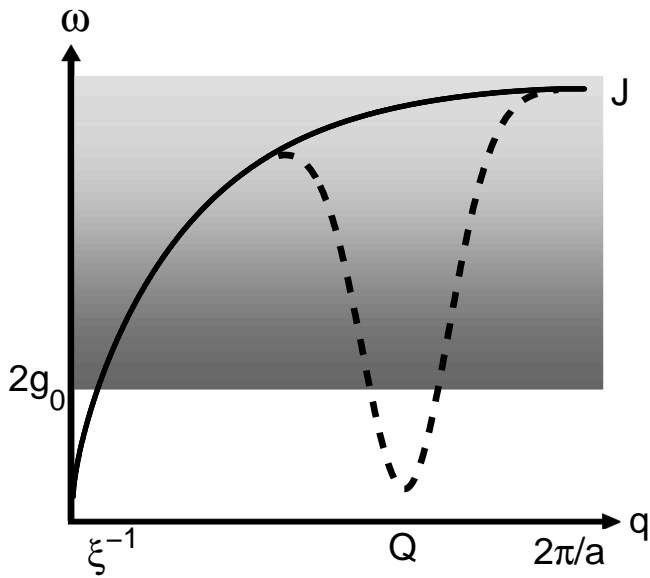}
\end{center}
\caption{\label{FIG_9}
Schematic dispersion relation for an itinerant electron system exhibiting an enhanced susceptibility at wavevector $Q$, leading to a low-temperature BCS-like ground state. The canonical spin wave dispersion (solid line) is steep and these modes contribute little to the free energy in the ordered state. The enhanced susceptibility at $Q$ leads to collective slow modes (dashed line) around this point, an effect which is enhanced by coupling of the spin to the charge and lattice degrees of freedom. The population of these collective modes with energy below the single-particle excitation gap $2g_0$ drives the thermal phase transition; above $T_C$ pseudogaps remain and $2g_0/T_C$ is larger than the canonical BCS value of 3.5.
}
\end{figure}

We are led to apparent contradictions between observed weak-coupling ground states and signatures of strongly coupled electrons. The existence of very different energy scales, even for a fairly weak coupling model, helps to resolve the paradox. In the weak-coupling formula, the gap scale is still set by the (large) Fermi energy multiplied by a small weak-coupling factor: $g \sim E_Fexp(-1/\lambda)$. In a pure BCS theory, the thermal transition would be produced solely by particle-hole fluctuations across the gap, which gives rise to the canonical relation $2g_0/T_C = 3.5$. Such a picture neglects the collective modes (spin-waves, or phase modes, and in a coupled system like Cr spin-phonon modes); see Fig.~\ref{FIG_9}. In a model with purely electronic interactions, the spin-wave velocity is steep (canonically $v_F/3$), thus entering the particle-hole continuum at a wave-vector corresponding to the inverse of the coherence length $\xi^{-1} \sim g_0/v_F << a^{-1}$, which is much smaller than an inverse lattice constant. With a dispersion that is so steep, the thermal occupation of these modes contributes little to the free energy at low temperatures. 
However, at larger momenta, the spectral weight in the collective modes lies on the scale of the interaction strength $j$ (here $j \sim 140 \; \mathrm{meV}$), which is much \textit{softer} than the weak coupling theory allows. Furthermore, in a system with substantial magneto-phonon coupling (as evidenced here by the CDW), this spectral weight mixes with phonons on a characteristic scale of the Debye frequency; these slow modes have frequencies usually well within the gap, set by pinning and the phonon mass. Unless the CDW/SDW gap is truly tiny, it is usually the case that the population of these short-wavelength modes will drive the phase transition; above $T_C$ pseudogaps remain and $2g_0/T_C$ is large. This picture (Fig.~\ref{FIG_9}) is generic and can be applied equally well to CDW and SDW systems; charged superconductors themselves are special because long-range Coulomb forces stiffen the phase mode into the conventional plasmon.

By juxtaposing unambiguous proof of a weak-coupling ground state with signatures of incipient magnetic order at high temperature we have argued that the N\'{e}el transition in Cr differs from the expectations of mean-field theory, and that the distinction between strongly and weakly coupled systems of itinerant electrons should be significantly blurred. At sufficiently high pressure and low temperature, quantum fluctuations will cut off the decades long exponential evolution of the SDW and CDW order parameters, and pose new questions about the relationships between spin and charge order and the relevant energy scales. 

\section{Acknowledgments}

We are grateful to G. Aeppli and E. Isaacs for enlightening discussions, and to Dr. Junfeng Wang for generously sharing his transport data on the blue bronzes. The work at the University of Chicago was supported by NSF Grant No. DMR-0534296. R.J. acknowledges support from an NSF Graduate Research Fellowship. We acknowledge technical support from Vitali Prakapenka at GeoSoilEnviroCARS (Sector 13), Advanced Photon Source (APS), Argonne National Laboraotry. GeoSoilEnviroCARS is supported by the National Science Foundation - Earth Sciences (EAR-0622171) and Department of Energy - Geosciences (DE-FG02-94ER14466). Use of APS is supported by the U.S. DOE-BES, under Contract No. NE-AC02-06CH11357.

\end{document}